\def\be{\begin{equation}}
\def\te{\end{equation}}

\def\bea{\begin{eqnarray}}

\def\tea{\end{eqnarray}}

\def\bo{{\raise.15ex\hbox{\large$\Box$}}}

\def\TH{{\raise.2ex\hbox{$\displaystyle \bigodot$}\mskip-4.7mu \llap H \;}}
\def\face{{\raise.2ex\hbox{$\displaystyle \bigodot$}\mskip-2.2mu \llap {$\ddot
        \smile$}}}

\def\Hat#1{\rlap{\kern.10em$\widehat{\phantom G}$}#1}
\def\HAt#1{\rlap{\kern.05em$\widehat{\phantom G}$}#1}

\def\cap#1{\rlap{\kern.1em$\widehat{\phantom{G\vrule height.8em}}$}#1{}}
\def\Cap#1{\rlap{\kern.05em$\widehat{\phantom{G\vrule height.8em}}$}#1{}}

\def\leftrightarrowfill{$\mathsurround=0pt \mathord\leftarrow \mkern-6mu
        \cleaders\hbox{$\mkern-2mu \mathord- \mkern-2mu$}\hfill
        \mkern-6mu \mathord\rightarrow$}
\def\overleftrightarrow#1{\vbox{\ialign{##\crcr
        \leftrightarrowfill\crcr\noalign{\kern-1pt\nointerlineskip}
        $\hfil\displaystyle{#1}\hfil$\crcr}}}

\def\frac#1#2{{\textstyle{#1\over\vphantom2\smash{\raise.20ex
        \hbox{$\scriptstyle{#2}$}}}}}

\def\sfrac#1#2{{\vphantom1\smash{\lower.5ex\hbox{\small$#1$}}\over
        \vphantom1\smash{\raise.4ex\hbox{\small$#2$}}}}
\def\bfrac#1#2{{\vphantom1\smash{\lower.5ex\hbox{$#1$}}\over
        \vphantom1\smash{\raise.3ex\hbox{$#2$}}}}
\def\afrac#1#2{{\vphantom1\smash{\lower.5ex\hbox{$#1$}}\over#2}}

\catcode`@=11
\def\underline#1{\relax\ifmmode\@@underline#1\else
        $\@@underline{\hbox{#1}}$\relax\fi}
\catcode`@=12

\def\nis{\nointerlineskip}
\def\Abar{\vbox{\nis\moveright.33em\vbox{
        \hrule width.35em height.04em}\nis\kern.05em\hbox{$A$}}{}}
\def\Dbar{\vbox{\nis\moveright.20em\vbox{
        \hrule width.50em height.04em}\nis\kern.05em\hbox{$D$}}{}}
\def\Gbar{\vbox{\nis\moveright.20em\vbox{
        \hrule width.50em height.04em}\nis\kern.05em\hbox{$G$}}{}}
\def\mbar{\vbox{\nis\moveright.15em\vbox{
        \hrule width.60em height.04em}\nis\kern.05em\hbox{$m$}}{}}
\def\Rbar{\vbox{\nis\moveright.20em\vbox{
        \hrule width.50em height.04em}\nis\kern.05em\hbox{$R$}}{}}
\def\Vbar{\vbox{\nis\moveright.05em\vbox{
        \hrule width.60em height.04em}\nis\kern.05em\hbox{$V$}}{}}
\def\Xbar{\vbox{\nis\moveright.20em\vbox{
        \hrule width.60em height.04em}\nis\kern.05em\hbox{$X$}}{}}
\def\thetabar{\vbox{\nis\moveright.15em\vbox{
        \hrule width.30em height.04em}\nis\kern.05em\hbox{$\theta$}}{}}
\def\Lambdabar{\vbox{\nis\moveright.25em\vbox{
        \hrule width.35em height.04em}\nis\kern.05em\hbox{${\mit\Lambda}$}}{}}
\def\Sigmabar{\vbox{\nis\moveright.25em\vbox{
        \hrule width.50em height.04em}\nis\kern.05em\hbox{${\mit\Sigma}$}}{}}
\def\phibar{\vbox{\nis\moveright.18em\vbox{
        \hrule width.40em height.04em}\nis\kern.05em\hbox{$\phi$}}{}}
\def\chibar{\vbox{\nis\moveright.12em\vbox{
        \hrule width.40em height.04em}\nis\kern.05em\hbox{$\chi$}}{}}
\def\psibar{\vbox{\nis\moveright.23em\vbox{
        \hrule width.40em height.04em}\nis\kern.05em\hbox{$\psi$}}{}}
\def\debar{\vbox{\nis\moveright.18em\vbox{
        \hrule width.35em height.04em}\nis\kern.05em\hbox{$\partial$}}{}}
\def\delbar{\vbox{\nis\moveright.10em\vbox{
        \hrule width.63em height.04em}\nis\kern.05em\hbox{$\nabla$}}{}}

\newskip\humongous \humongous=0pt plus 1000pt minus 1000pt

\newif\ifdtup

\def\be{\begin{equation}}
\def\te{\end{equation}}
\def\bea{\begin{eqnarray}}
\def\tea{\end{eqnarray}}

\documentstyle[12pt]{article}

\makeatletter                                                           

\def\section{\@startsection {section}{1}{\z@}{-1.5ex plus -.5ex
minus -.2ex}{1ex plus .2ex}{\large\bf}}                                 

\def\@thmcountersep{}                                                   

\long\def\@makecaption#1#2{\vskip 10pt
\setbox\@tempboxa\hbox{#1. #2}   
   \ifdim \wd\@tempboxa >\hsize   
       #1. #2\par                 
     \else                        
       \hbox to\hsize{\hfil\box\@tempboxa\hfil}                         
   \fi}                                                                 

\def\ps@headings{                                                       
 \def\@oddhead{\footnotesize\rm\hfill\runninghead\hfill}
 \def\@evenhead{\@oddhead}                                              
 \def\@oddfoot{\rm\hfill\thepage\hfill}\def\@evenfoot{\@oddfoot} }

\makeatother                                                            

\begin{document}

\title{Decoherence of Correlation Histories}{}{}

\def\runninghead{CALZETTA AND HU: DECOHERENCE OF CORRELATION HISTORIES}
\author {
{\em Esteban Calzetta} \thanks
{IAFE, cc 167, suc 28, (1428) Buenos Aires, Argentina}
\and
{\em B. L. Hu} \thanks
{Department of Physics, University of Maryland,  College Park, MD 20742, USA}}

\date{} 

\pagestyle{headings}                                                   
\flushbottom                                                           

\maketitle
\vspace{-10pt} 

\begin{abstract}
We use a $\lambda\Phi^4$ scalar quantum field theory to illustrate
a new approach to the study of quantum to classical transition.
In this approach, the decoherence functional is employed to assign
probabilities to consistent histories defined in terms of
correlations among the fields at separate points, rather than the
field itself. We present expressions for the quantum amplitudes
associated with such histories, as well as for the decoherence
functional between two of them. The dynamics of an individual
consistent history may be described by a Langevin-type equation,
which we derive.

\end{abstract}

\noindent {\it Dedicated to Professor Brill on the occasion of his sixtieth
birthday, August 1993}


\section{Introduction}

\subsection{Interpretations of Quantum Mechanics and Paradigms of Statistical
Mechanics}

This paper attempts to bring together two basic concepts, one from the
foundations
of statistical mechanics and the other from the foundations of quantum
mechanics,
for the purpose of addressing two basic issues in physics:\\
1) the quantum to classical transition, and\\
2) the quantum origin of stochastic dynamics.\\
Both issues draw in the interlaced effects of dissipation, decoherence,
noise, and fluctuation. A central concern is the role played by coarse-graining
--the naturalness of its choice, the effectiveness of its implementation
and the relevance of its consequences.

On the fundations of quantum mechanics, a number of alternative
interpretations exists, e.g., the Copenhagen interpretation,
the many-world interpretation \cite{manyworld}, the consistent history
interpretations \cite{conshis}, to name just a few (see \cite{HarMis}
for a recent review).
The one which has attracted
much recent attention is the decoherent history approach of Gell-Mann and
Hartle \cite{decohis}. In this formalism,
the evolution of a physical system is described in terms of
`histories': A given history may be either exhaustive (defining
a complete set of observables at each instant of time) or
coarse-grained. While in classical physics each history is assigned
a given probability, in quantum physics a consistent assignment
of probabilities is precluded by the overlap between different
histories. The decoherence functional gives a quantitative
measure of this overlap; thus the quantum to classical
transition can be studied as a process of ``diagonalization''
of the decoherence  functional in the space of histories.

On the foundational aspects of statistical mechanics, two major paradigms
are often used to describe non-equilibrium processes (see, e.g.,
\cite{Akhiezer,Prigogine,KadBay,KuboTH}):
the Boltzmann theory of molecular kinetics,
and the Langevin (Einstein-Smoluchowski) theory of
Brownian motions. The difference between the two are of both formal and
conceptual character.

To begin with, the {\em setup} of the problem
is different:
In kinetic theory one studies the overall dynamics of a system of gas
molecules, treating each molecule in the system on the same footing,
while in Brownian motion one (Brownian) particle which defines the system is
distinct from the rest, which is relegated as the environment. The terminology
of `revelant' versus `irrevelant' variables highlights the discrepancy.

The {\em object} of interest in kinetic theory is the (one-particle)
distribution function (or the nth-order correlation function), while
in Brownian motion it is the reduced density matrix. The emphasis in
the former is
the correlation amongst the particles, while in the latter is
the effect of the environment on the system.

The nature of {\it coarse-graining} is also very different: in kinetic theory
coarse-graining resides in the adoption of the molecular chaos assumption
corresponding formally to a truncation of the BBGKY hierarchy, while in
Brownian motion it is in the integration over the environmental variables.
The part that is truncated or `ignored' is what constitutes the noise,
whose effect on the `system' is to introduce dissipation in its dynamics.
Thus the fluctuation-dissipation relation and other features.

Finally the {\it philosophy} behind these two paradigms are quite different:
In Brownian motion problems, the separation of the system
from the environment is prescribed: it is usually determined by some
clear disparity
between the two systems. These models represent ``autocratic systems'',
where some degrees of freedom are more relevant than others.
In the lack of such clear distinctions,
making a separation `by hand' may seem rather {\it ad hoc} and unsatisfactory.
By contrast, models subscribing to the kinetic theory paradigm represent
``democratic systems'': all particles in a gas are equally relevant.
Coarse-graining in Boltzmann's kinetic theory appears less contrived,
because information about higher correlation orders usually reflects the
degree of precision in a measurement, which is objectively definable.

In the last five years we have explored these two basic paradigms of
non-equilibrium statistical mechanics in the framework of
interacting quantum field theory with the aim of treating
dissipative processes in the early universe \cite{CalHu88,CHH88}
and decoherence processes in the quantum to classical transition issue
\cite{ZhangPhD}.
Here we have begun to explore the issues of decoherence with the kinetic model.

Because of the difference in approach and emphasis between these two paradigms
and in view of their fundamental character,
it is of interest to build a bridge between them. We have recently
carried out such a study with quantum fields \cite{CalHuKT2BM}.
By delineating the conditions
under which the Boltzmann theory reduces to the Langevin theory,
we sought answers to the following questions:\\
1) What are the factors condusive to the evolution of a `democratic system'
to an `autocratic system' and vise versa ? A  more
natural set of criteria for the separation of the system from the environment
may arise from the interaction and dynamics of the initial closed system
\cite{HuSpain,HuWaseda}.\\
2) The construction of collective variables from the basic variables,
the description of the dynamics of the collective variables, and the depiction
of the behavior of a
coarser level of structure emergent from the microstructures.
\cite{HuSpain,HuErice,BalVen}.

The paradigm of quantum open systems described by quantum Brownian models
has been used to analyze the decoherence and dissipation processes,
for addressing basic issues like quantum to classical transitions,
fluctuation and noise, particle creation and backreaction, which arise in
quantum
measurement theory \cite{Zurek,Zeh,WheZur}, macroscopic quantum systems
\cite{CalLeg83},
quantum cosmology \cite{HuTsukuba} (for earlier
work see references in \cite{HalGR13}),
semiclassical gravity
\cite{PazSin1,HuPhysica,CalCQG}, and inflationary cosmology
\cite{HuBelgium}.
The reader is referred to these references and references
therein for a description of this line of study.

The aim of this paper is to explore the feasibility for addressing the same
set of basic issues using the kinetic theory paradigm.
We develop a new approach based on
the application of the decoherence functional
\cite{conshis,decohis} formalism to
histories defined in terms of {\em correlations} between the
fundamental field variables.
We shall analyse the decoherence between different histories
of an interacting quantum field, a $\lambda\Phi^4$ theory here taken as
example,
corresponding to different particle spectra and study issues on
the physics of quantum to classical transition, the
relation of decoherence to dissipation, noise and fluctuation,
and the quantum origin of classical stochastic dynamics.

\subsection{Quantum to Classical Transition and Coarse-Graining}

One basic constraint in the building of quantum theory is that it
should reproduce classical mechanics in some limit.
(For a schematic discussion of the different criteria of classicality
and their relations, see \cite{HuZhaDrexel}).
Classical behavior can be characterized
by the existence of strong correlations between position and
momentum variables described by the classical equations of motion
\cite{correlation}
and by the absense of interference phenomena (decoherence).

Recent research in quantum gravity and cosmology have focussed on the
issue of quantum to classical transition. This was highlighted by
quantum measurement theory for closed systems
(for a general discussion, see, e.g., \cite{HarBri}),
the intrinsic incompatibility of
quantum physics with general relativity \cite{Penrose},
and the quantum origin of classical fluctuations in explaining the large scale
structure of the Universe.
Indeed, in the inflationary models of the Universe \cite{inflation},
one hopes to trace
all cosmic structures to the evolution from quantum perturbations in the
inflaton field. More dramatically, in quantum cosmology
\cite{QC} the whole (classical) Universe where we now live in
is regarded as the outcome of a quantum to classical transition on a cosmic
scale.  In these models, one hopes not only to explain the `beginning' of
the universe as a quantum phenomenon, but also to account for the classical
features of the present universe as a consequence of quantum fluctuations.
This requires not only a theoretical understanding of the quantum to
classical transition issue in quantum mechanics,
but also a theoretical derivation of the laws of classical stochastic mechanics
from quantum mechanics,
the determination of the statistical properties of classical noise
(e.g., whether it is white or coloured, local or nonlocal) being an essential
step in the formulation of a microscopic theory of the structure of the
Universe
\cite{HuBelgium}.

Our understanding of the issue of quantum to classical transition has been
greatly advanced by the recent development of the decoherent
histories approach to quantum mechanics \cite{decohis}.
An essential element of the decoherent histories approach is
that the overlap between two exhaustive histories can never
vanish. Therefore, the discussion of a quantum to classical
transition can only take place in the framework of a coarse
grained description of the system, that is, giving up
a complete specification of the state of the system at any
instant of time.

As a matter of fact, some form of coarse graining underlies
most, if  not all, successful macroscopic physical theories.
This fact has been clearly recognized and exploited at least
since the work of Nakajima and Zwanzig
\cite{Nakajima,Prigogine} on the
foundations of nonequilibrium statistical mechanics. Like
statistical mechanics, the decoherent histories approach
allows a variety of coarse-graining procedures; not all of
these, however, are expected to be equally successful in leading
to interesting theories. Since the prescription of the coarse
graining procedure is an integral part of the implementation
of the decoherent histories approach, the development and
evaluation of different coarse graining strategies is
fundamental to this research program.

When we survey the range of meaningful macroscopic (effective) theories in
physics arising from successfully coarse-graining a microscopic (fundamental)
theory, one particular class of examples is outstanding;
namely, the derivation of the hydrodynamical description
of dilute gases from classical mechanics.
The crucial step in deriving the Navier-Stokes
equation for a dilute gas consists in rewriting the Liouville
equation for the classical distribution function as a
BBGKY hierarchy, which is then truncated by invoking a
`molecular chaos' assumption. If the truncation is made
at the level of the two-particle reduced distribution function,
the Boltzmann equation results. In the near-equilibrium limit,
this equation leads to the familiar Navier-Stokes theory.

We must stress that in this general class of theories exemplified by
Boltzmann's
work,  coarse-graining is  introduced through the truncation of the
hierarchy of distribution functions; i.e.,
by neglecting correlations of some order and above at some singled-out time
\cite{Akhiezer,Prigogine}.  This type of coarse graining strategy is
qualitatively different from those  used in most of the recent work in
quantum measurement theory and cosmology, which invoke a system-bath,
space-time, or momentum-space separation. In most of these cases,
an intrinsically justifiable division of the system from the environment
is lacking and one has to rely on case-by-case physical rationales
for making such splits.
(An example of system-bath split is Zurek's description of the
measurement process in quantum mechanics, where a bath is explicitly
included to cause decoherence in the system-apparatus complex \cite{Zurek}.
Space-time coarse graining has been discussed by Hartle \cite{HarSTCG} and
Halliwell {\it et al} \cite{HalSOH}. An example of coarse-graining in
momentum space
is stochastic inflation  \cite{StoInf}, where inflaton modes
with wavelenghts shorter than the horizon are treated as an environment
for the longer wavelenght modes \cite{HuZha90}).

\subsection{Coarse-Graining in the Hierarchy of Correlations}

In this paper we shall develop a version of the decoherent
histories approach where the coarse-graining procedure is patterned
after the truncation of the BBGKY hierarchy of distribution functions.
For simplicity, we shall refer below to the theory
of a single scalar quantum field, with a $\lambda\Phi^4$- type nonlinearity.

The simplest quantum field theoretical analog to
the hierarchy of distribution functions in statistical mechanics
is the sequence of Green functions (that is, the expectation values
of products of $n$ fields) \cite{CalHu88}. In this
approach, the BBGKY hierarchy of kinetic equations is replaced
by the chain of Dyson equations, linking each Green function
to other functions of higher order.

The analogy between these two hierarchies is rendered most
evident if we introduce ``distribution functions'' in field
theory through suitable partial Fourier transformation of
the Green functions. Thus, a ``Wigner function'' \cite{Wigner} may be
introduced as the Fourier transform of the Hadamard function
(the symmetric expectation value of the product of two
fields) with respect to the difference between its arguments.
It obeys both a mass shell constraint
and a kinetic equation, and may be regarded as
the physical distribution function for a gas of
quasi particles, each built out of a cloud of virtual quanta.
Similar constructs may be used to
introduce higher ``distribution functions'' \cite{CalHu88,KadBay}.

As in statistical mechanics, the part of a given Green function
which cannot be
reduced to products of lower functions defines the corresponding
``correlation function''. Thus the chain of Green functions
is also a hierarchy of correlations.

To establish contact between the hierarchy of Green functions
and the decoherent histories approach, let us recall the well-known
fact that the set of expectation values of all field
products contains in itself all the information about the
statistical state of the field \cite{CalHu88}. For a scalar field theory
with no symmetry breaking, we can even narrow this set to
products of even numbers of fields. This result suggests
that a history can be described in terms of the values of suitable
composite operators, rather than those of the fundamental
field.  If products to all orders are
specified (binary, quartet, sextet, etc), then the
description of the history is exhaustive, and different
histories do not decohere. On the other hand, when some
products are not specified, or when the information of higher
correlations are missing, which is often the case in realistic measurement
settings, the description is coarse-grained, which can lead to decoherence.

In this work we shall consider coarse-grained
histories where the lower field products (binary, quartic)
are specified, and higher products are not. Decoherence will
mean that the specified composite operators can be assigned definite
values with consistent probabilities. Higher composite operators
retain their quantum nature, and therefore cannot be assigned
definite values. However, their expectation values can be
expressed as functionals of the specified correlations by
solving the corresponding Dyson equations with suitable
boundary conditions. This situation is exactly analogous
to that arising from the truncated BBGKY hierarchy, where the
molecular chaos assumption allows the expression of higher
distribution functions as functionals of lower ones ( e.g.,
\cite{Akhiezer,Prigogine}).

For those products of fields which assume definite values with
consistent probabilities,
these values can be introduced as stochastic variables in the dynamical
equations for the other quantities of interest (usually of lower
correlation order). This approach would provide a theoretical basis for
the derivation of the equations of classical stochastic dynamics from
quantum fields. It can offer a justification (or refutation)
for a procedure commonly assumed but never proven in some popular theories
like stochastic inflation \cite{StoInf,HuZha90}.
Moreover, since in general we shall obtain nontrivial ranges of values for
the specified products with
nonvanishing probabilities, it can be said that our procedure
captures both the average values of the field products and the
fluctuations around this average. The statistical nature of these
fluctuations is a subject of great interest in itself \cite{HuBelgium}.

There is another conceptual issue that our approach may
help to clarify.
As we have already noted, in the system-bath split approach
to coarse-graining, as well as in related procedures, it
is crucial to introduce a hierarchical order among the
degrees of freedom of the system, in such a way that some
of them may be considered relevant, and others irrelevant.
While it is often the case that the application itself
suggests which notions of relevance may lead to an
interesting theory, in a quantum cosmological model,
which purports to be a ``first principles'' description of
our Universe, all these choices are, in greater or lesser
degree, arbitrary. Since correlation functions already
have a ``natural'' built-in hierarchical ordering, in this approach
the `arbitrariness' is reduced to deciding on which level this hierarchy
is truncated, and that in turn is determined by the degree of precision one
carries out the measurement. In most case  one still needs to show
the robustness of the macroscopic result against the variance of the extent
of coarse-graining, and exceptional situations do exists (an example is
the long time-tail relaxation behavior in multiple particle scattering of
dense  gas,    arising  from  a  failure  of  the  simple  molecular  chaos
assumption).
But in general terms correlational coarse-graining seems to us a less
{\it ad hoc}
procedure compared to the commonly used system-bath splitting and
coarse-graining.

This paper is organized as follows:  In Sec. 2
we discuss the implementation of our
procedure for the simple case of a $\lambda\Phi^4$ theory
in flat space time. We then derive the formulae for the quantum
amplitude associated with a set of correlation histories and  the decoherence
functional between two such histories. In Sec. 3 we discuss the decoherence
of correlation histories between binary histories and derive the classical
stochastic source describing the effect of higher-order
correlations on the lower-order ones, arriving at a Langevin equation for
classical stochastic dynamics.
In Sec. 4 we summarize our findings.

\section{Quantum Amplitudes for Correlation Histories and Effective Action}

\subsection{Quantum Mechanical Amplitudes for Correlation Histories}

In this section, we shall consider
the quantum mechanical amplitudes associated with different
histories for a $\lambda\Phi^4$ quantum field theory, defined
in terms of the values of time-ordered products of
even numbers of fields at various space time points.
Let us begin by motivating our ansatz for the
amplitudes of these correlation histories.

In the conceptual framework of decoherent histories \cite{decohis},
the ``natural'' exhaustive specification of a
history would be to define the value of the field $\Phi (x)$
at every space time point. These field values are c numbers.
The quantum mechanical amplitude for a given history is
$\Psi [\Phi ]\sim e^{iS[\Phi ]}$, where $S$ is the classical
action.
The decoherence functional between two different
specifications is given by $D[\Phi ,\Phi ']\sim
\Psi [\Phi ]\Psi [\Phi ']^*$. Since $\vert D[\Phi ,\Phi ']\vert
\equiv 1$, there is never decoherence between these histories.

A coarse-grained history would be defined in general through
a ``filter function'' $\alpha$, which is
basically a Dirac $\delta$ function concentrated on
the set of exhaustive histories matching the specifications
of the coarse-grained history. For
example, we may have a system with two degrees of freedom
$x$ and $y$, and define a coarse-grained history by specifying
the values $x_0(t)$ of $x$ at all times. Then the filter
function is $\alpha [x,y]=\prod_{t\in R}\delta (x(t)-x_0(t))$.
The quantum mechanical amplitude for the coarse-grained
history is defined as

\begin{equation}
\Psi [\alpha ]=\int~D\Phi~e^{iS}\alpha [\Phi ]\label{psiofalfa}
\end{equation}
where the information on the quantum state of the field is assumed
to have been included in the measure and/or the boundary
conditions for the functional integral. The decoherence
functional  for two coarse-grained histories is \cite{decohis}

\begin{equation}
D[\alpha, \alpha ']=\int~D\Phi D\Phi ' e^{i(S(\Phi )-S(\Phi '))}
\alpha [\Phi ]\alpha '[\Phi ']\label{dofalfa}
\end{equation}

In this path integral expression, the two histories $\Phi$ and
$\Phi '$ are not independent; they assume identical values on a
$t=T={\rm constant}$ surface in the far future. Thus, they may be
thought of as a single, continuous history defined on a two-branched
``closed time-path'' \cite{SchKel,Zhou,DeWJor,87},
the first branch going from
$t=-\infty$ to $T$, the second from $T$ back to $-\infty$.
Alternatively, we can think of $\Phi=\Phi^1$ and $\Phi '=\Phi^2$
as the two components of a field doublet defined on ordinary space
time  \cite{CalHu88}, whose classical action is
$S[\Phi^a]=S[\Phi^1]-S[\Phi^2]$.
This notation shall be useful later on.

Let us try to generalize this formalism to correlation histories.
We begin with the simplest case, where only binary products are specified.
In this case a history is defined by identifying a symmetric kernel
$G(x,x')$, which purports to be the value of the product
$\Phi (x)\Phi (x')$ in the given history,
both $x$ and $x'$ defined in  Minkowsky space - time. By analogy
with the formulation above, one would write the quantum mechanical
amplitude for this correlation history as

\begin{equation}
\Psi [G]=\int~D\Phi~e^{iS}\prod_{x\gg x'}\delta (\Phi (x)\Phi
(x')-G(x,x'))\label{psiofG}
\end{equation}
(In this equation, we have introduced a formal ordering of points
in Minkowsky space - time, simply to avoid counting the same
pair twice.)

But this straightforward generalization for the correlation history
amplitude is unsatisfactory
on at least two counts. First, it assumes that the given kernel
$G$ can actually be decomposed (maybe not uniquely) as a product
of c number real fields at different locations; however, we wish to
define amplitudes for kernels (such as the Feynman propagator)
which do not have this property. Second, (which is related to the first
point,) it is ambiguous, since we do not have a unique way to
express higher even products of fields in terms of binary
products, and thus of applying the $\delta$ function constraint.

To give an example of this, observe that, should we
expand the exponential of the action in powers of
the coupling constant $\lambda$, the
second order term
$\int~dx~dx'~\Phi (x)^4\Phi (x')^4$
could become, after integration over the delta function, either
\begin{eqnarray}
\int~dx~dx'&&~G(x,x)^2G(x',x')^2,\nonumber\\
\int~dx~dx'&&~G(x,x')^4,\nonumber\\
\end{eqnarray}
or any other combination; of course, if $G$ could be decomposed
as a product of fields, this would be unimportant.

Let us improve on these shortcomings. The general idea is
to accept Eq. (\ref{psiofG}) as the definition of the amplitude in the
restricted set of kernels where it can be applied, and to define
the amplitude for more general kernels through some process
of analytical continuation. To this end, we must rewrite the
quantum mechanical amplitude in a more transparent form,
which we achieve by using an integral representation of the
$\delta$ function. Concretely, we redefine

\begin{equation}
\Psi [G]=\int~DK\int~D\Phi~e^{iS+{i\over 2}\int dxdx'~K(x,x')(\Phi (x)
\Phi (x')-G(x,x'))}\label{psiofG,K}
\end{equation}
where the filter function in the Gell-Mann Hartle scheme is replaced by
an integration
over ``all'' symmetric non-local sources $K$.
Eq. (\ref{psiofG,K}) is not yet a complete definition,
since one must still specify both the path and the measure to be used
in the $K$ integration.
Performing the integration over fields, we obtain

\begin{equation}
\Psi [G]=\int~DK~e^{i(W[K]-(1/2)KG)}\label{psiofG,W}
\end{equation}
where $W[K]$ is the generating functional for connected vacuum
graphs with $\lambda\Phi^4$ interaction, and
$(\Delta^{-1}-K)^{-1}$
for propagator (see below). Here
$\Delta^{-1}=-\nabla^2+m^2$ is the free propagator for our scalar field
theory (our sign convention for the flat space - time metric is $-+++$).

The path integral over kernels can be computed
through functional techniques. For example, for a free field,
$\lambda =0$,

\begin{equation}
W[K]=-i\ln {\rm Det} [(\Delta^{-1}-K)^{-1/2}]+{\rm constant}\label{WofK}
\end{equation}
Through the change of variables

\begin{equation}
(\Delta^{-1}-K)=\kappa G^{-1}\label{varchange}
\end{equation}
 we obtain

\begin{equation}
\Psi[G]=~{\rm constant}~[{\rm Det~G}]^{-1/2}~e^{(-i/2)\Delta^{-1}G}
\label{freepsi}
\end{equation}

When the self coupling
$\lambda$ is not zero, the evaluation of $\Psi[G]$ is
more involved; however, if we are interested in the leading
behavior of the amplitude only, we can simply evaluate the
functional integral over $K$ by saddle point methods. The saddle
lies at the solution to

\begin{equation}
{\partial W[K]\over\partial K}={1\over 2}G\label{saddle}
\end{equation}
We recognize immediately that the exponent, evaluated at the saddle
point, is simply the 2 Particle Irreducible (2PI) effective action
$\Gamma$, with $G$ as propagator (see below).
Including also the integration on gaussian fluctuations around the
saddle, we find

\begin{equation}
\Psi [G]\sim [{\rm Det}\{{\partial^2\Gamma\over\partial G^2}\}]^{(1/2)}
e^{i\Gamma [G]}\label{psiofG,2PI}
\end{equation}

This is our main result.

As a check, it is interesting to compare the
saddle method expression with our exact result for free fields.
For a free field
$\Gamma [G]=(-i/2)\ln {\rm Det} (G)~-~(1/2)\Delta^{-1}G$,
and therefore
$\Gamma_{,G}=(-i/2)(G^{-1}-i\Delta^{-1})$,
$\Gamma_{,G,G}=(i/2)G^{-2}$, so

\begin{equation}
[{\rm Det}\{{\partial^2\Gamma\over\partial G^2}\}]^{(1/2)}
e^{i\Gamma [G]}=[{\rm Det} G]^{-1}[{\rm Det} G]^{1/2}
e^{(-i/2)\Delta^{-1}G}
\end{equation}
which is exactly the earlier result, Eq. (\ref{freepsi}).

\subsection{Quantum Amplitudes and Effective Actions}

Eq. (\ref{psiofG,2PI}) is the natural generalization to
correlation histories of the quantum mechanical amplitude $e^{iS}$
associated to a field configuration.
Let us consider its physical meaning.

The effective action is usually introduced in Field Theory
books \cite{Ramond} as a compact device to generate the Feynman graphs
of a given theory. Indeed, all Feynman graphs appear in the
expansion  of the generating functional

\begin{equation}
Z[J]=\int~D\Phi e^{i(S+J\Phi )}\label{Z[J]}
\end{equation}
in powers of the external source $J$ [here, $J\Phi =\int~d^4x~J(x)
\Phi (x)$]. $Z$ has the physical meaning of a vacuum persistance
amplitude: it is the amplitude for the in vacuum (that is, the
vacuum in the distant past) to evolve into the out vacuum (the
vacuum in the far future) under the effect of the source $J$. Thus,
after proper normalization,  $\vert Z\vert$ will be
unity when the source is unable to create pairs out of the vacuum,
and less than unity otherwise.

A more compact representation of the Feynman graphs is provided
by the functional $W[J]=-i\ln Z[J]$; the Taylor expansion of $W$
contains only connected Feynman graphs.
Thus $W$ developing a (positive) imaginary part signals
the instability of the vacuum under the external source $J$.

The external source will generally drive the quantum field $\Phi$
so that its matrix element

\begin{equation}
\phi (x)={\langle 0out\vert\Phi (x)\vert 0in\rangle
\over\langle 0out\vert 0in\rangle}\label{VEV}
\end{equation}
 between the in and out vacuum states will not be zero.
Indeed, it is easy to see that

\begin{equation}
\phi ={\partial W\over\partial J}\label{VEVpart}
\end{equation}

The transformation from $J$ to $\phi$ is generally one to one,
and thus it is possible to consider the matrix element, and not
the source, as the independent variable. This is achieved by
submitting $W$ to a Legendre transformation, yielding the
effective action $\Gamma [\phi ]=W[J]-J\phi$
[$J$ and $\phi$ being related through Eq. (\ref{VEVpart})]. This equation
can be inverted to yield the dynamic law for $\phi$

\begin{equation}
{\partial\Gamma\over\partial\phi}=-J\label{VEVdyn}
\end{equation}
Eq. (\ref{VEVdyn}) shows that $\Gamma$ may be thought of as a
generalization of the classical action, now including quantum
effects. In the absence of external sources, the in and out
vacua agree, so $\phi$ becomes a true expectation value; its
particular value is found by extremizing the effective action.
Indeed, in this case it can be shown that $\Gamma$ is the
energy of the vacuum.

$\Gamma [\phi ]$ can be defined independently of the external
source through the formula \cite{Jackiw}

\begin{equation}
\Gamma [\phi ]=S[\phi ]+(i/2)\ln {\rm Det}
{}~({\partial^2S\over\partial\phi^2})+\Gamma_1[\phi ]\label{1PIEA}
\end{equation}
where $\Gamma_1$ represents the sum of all one particle irreducible (1PI)
vacuum graphs of an auxiliary theory whose classical action is
obtained from expanding the classical action $S[\phi +\varphi]$ in
powers of $\varphi$, and deleting the constant and linear terms. Eq.
(\ref{1PIEA}) shows that $\Gamma$ is related to the vacuum persistance
amplitude of quantum fluctuations around  the matrix element $\phi$.
Therefore, an imaginary part in $\Gamma$ also signals a vacuum instability.
This situation closely resembles the usual approach to tunneling and
phase transitions,
where an imaginary part in the free energy signals the onset of
instability \cite{Langer}.

Observe that each of the transformations from $Z$ to $W$ to $\Gamma$
entails a drastic simplification of the corresponding Feynman graphs
expansions, from all graphs in $Z$ to connected ones in $W$ and to
1PI ones in $\Gamma$. Roughly speaking, it is unneccessary to
include non 1PI graphs in the effective action, because the sum of
all one-particle insertions is already prescribed to add up to
$\phi$. Now the process can be continued: if we could fix in
advance the sum of all self energy parts, then we could write down
a perturbative expansion where only 2PI Feynman graphs need be
considered.  This is achieved by the 2PI effective action \cite{2PI}.

Let us return to Eq. (\ref{Z[J]}), and add to the external source
a space-time dependent mass term

\begin{equation}
Z[J,K]=\int~D\Phi e^{i(S+J\Phi +(1/2)\Phi K\Phi )}\label{Z[J,K]}
\end{equation}
where $\Phi K\Phi =\int~dx~dx'~\Phi (x)K(x,x')\Phi (x')$. Also
define $W[J,K]=-i\ln Z[J,K]$. Then the variation of $W$ with
respect to $J$ defines the in-out matrix element of the
field, as before, but now we also have

\begin{equation}
{\partial W\over\partial K(x,x')}={1\over 2}[\phi (x)\phi (x')
+G_F(x,x')]\label{dWdK}
\end{equation}
where $G_F$ represents the Feynman propagator of the
quantum fluctuations $\varphi$ around the matrix element $\phi$.
As before, it is possible to adopt $G$ as the independent variable,
instead of $K$. To do this, we define the 2PI effective action
(in schematic notation)
$\Gamma [\phi ,G_F]=W[J,K]-J\phi -(1/2)K[\phi^2+G_F]$.
Variation of this new $\Gamma$ yields the equations of motion
$\Gamma_{,\phi}=-J-K\phi$, $\Gamma_{,G_F}=(-1/2)K$.

We can see that the 2PI effective action generates the dynamics
of the Feynman propagator, and in this sense it plays for it
the role that the classical action plays for the field.
In this sense we can say that Eq. (\ref{psiofG,2PI})
generalizes the usual definition of quantum mechanical amplitudes.

The perturbative expansion of the 2PI effective action reads \cite{2PI}

\begin{equation}
\Gamma [\phi ,G_F]=S[\phi ]+(i/2)\ln {\rm Det} G_F^{-1}
+({1\over 2})~{\rm Tr}({\partial^2S\over\partial\phi^2}G_F)
+\Gamma_2[\phi ,G_F]+{\rm constant}\label{2PIEA}
\end{equation}
where $\Gamma_2$ is the sum of all 2PI vacuum graphs of the
auxiliary theory
already considered, but with $G_F$ as propagator in the internal lines.
As we anticipated, to replace $G_F$ for the perturbative
propagator amounts to adding all self energy insertions, and
therefore no 2PI graph needs  be explicitly included.

Like its 1PI predecessor, the 2PI effective action has the physical meaning
of a vacuum persistence amplitude for quantum fluctuations $\varphi$,
constrained to have vanishing expectation value and a given
Feynman propagator. Therefore, an imaginary part in the 2PI
effective action also signals vacuum instability.

The description of the dynamics of a quantum field through
both $\phi$ and $G_F$ simultaneously, rather than $\phi$
alone, is appealing not only because it allows one to perform
with little effort the resummation of an infinite set of
Feynman graphs, but also because for certain quantum states,
it is possible to convey statistical information about the field
through the nonlocal source $K$. This information is
subsequently transferred to the propagator. For this reason,
the 2PI effective action formalism is, in our opinion, a most suitable tool
to study statistical effects in field theory, particularly for
out-of-equilibrium fields \cite{CalHu88,Spino}.
In our earlier studies the object of interest
is the on-shell effective action, that is, the effective
action for propagators satisfying the equations of motion.
Here, in Eq. (\ref{psiofG,2PI}), we find a relationship between
the quantum mechanical amplitude for a correlation
history and the 2PI effective action which does not
assume any restriction on the propagator concerned.

\subsection{Quantum Amplitudes for More General Correlation Histories:
2PI CTP Effective Action}
One of the peculiarities of the ansatz Eq.(\ref{psiofG}) for
the amplitude of a correlation history is that the kernel
$G$ must be interpreted as a time - ordered binary product
of fields. This results from the known feature of the path
integral, which automatically time orders any monomials
occurring within it. Before we proceed to introduce the
decoherence functional for correlation histories, it is
convenient to discuss how this restriction could be lifted,
as well as the restriction to binary products.

The time ordering feature of the path integral is also
responsible for the fact that the c-number field $\phi$
in Sec. 2.2 is a matrix element, rather than a true
expectation value. As a matter of fact, the Feynman propagator
$G_F$ discussed in the previous section is also a
matrix element

\begin{equation}
G_F (x,x')={\langle 0out\vert T[\varphi (x)\varphi (x')]\vert 0in\rangle
\over\langle 0out\vert 0in\rangle}\label{GF}
\end{equation}
Because $\phi$ and $G_F$ satisfy mixed boundary conditions,
the dynamic equations resulting from the 2PI effective action
are generally not causal. This drawback has placed limitations in their
physical applications.

Schwinger \cite{SchKel} has introduced an extended effective action, whose
arguments are true expectation values with respect to some in
quantum state. Because the dynamics of these expectation values
may be formulated as an initial value problem, the equations
of motion resulting from the Schwinger-Keldysh effective action are causal.
Schwinger's idea is also the key to solving the restrictions in
our definition of quantum amplitudes for correlation histories.

Schwinger's insight was to apply the functional formalism we
reviewed in Sec. {\bf 2.2} to fields defined on a ``closed time-path'',
composed of a ``direct'' branch $-T\le t\le T$,
and a ``return'' branch $T\ge t\ge -T$ (with $T\to\infty$)
\cite{SchKel,Zhou}.
Actually, we have already encountered this kind of path in the
discussion of the decoherence functional for  coarse - grained
histories. Since the path doubles back on itself, the in vacuum
is the physical vacuum at both ends; the formalism may be
generalized to include more general initial states, but we shall
not discuss this possibility\cite{CalHu88}.

The closed time-path integral time-orders products of fields
on the direct branch, anti-time-orders fields on the
return branch, and places fields on the return branch always
to the left of fields in the direct branch. To define the
closed time-path generating functional, we must introduce
two local sources $J_a$, and four nonlocal ones $K_{ab}$
(as in Sec. {\bf 2.1} an index $a,b=1$ denotes a point on the first
branch, while an index $2$ denotes a point on the return part of the path).
These sources are conjugated to c number fields $\phi^a$
and propagators $G^{ab}$, which stand for
$\langle 0in\vert \Phi^a (x)\vert 0in\rangle$ and
$\langle 0in\vert \varphi^a (x)\varphi^b (x')]\vert 0in\rangle$.
Explicitly, decoding the indices, the
propagators are defined as (here and from now on, we assume that
the background fields $\phi^a$ vanish):

\begin{equation}
G^{11}(x,x')=\langle 0in\vert T[\Phi (x)\Phi (x')]\vert 0in\rangle
\label{Dick}
\end{equation}

\begin{equation}
G^{12}(x,x')=\langle 0in\vert \Phi (x')\Phi (x)\vert 0in\rangle
\label{negative}
\end{equation}

\begin{equation}
G^{21}(x,x')=\langle 0in\vert \Phi (x)\Phi (x')\vert 0in\rangle
\label{positive}
\end{equation}

\begin{equation}
G^{22}(x,x')=\langle 0in\vert (T[\Phi (x)\Phi (x')])^{\dagger}\vert 0in\rangle
\label{dyson}
\end{equation}

They are, respectively, the Feynman, negative- and positive-
frequency Wightman, and Dyson propagators. The definition of the closed
time-path (CTP) or in-in 2PI effective action follows the
same steps as the ordinary effective action discussed in the
previous section, except that now, besides space-time integrations, one must
sum over the discrete indexes $a,b$. These indexes can be raised
and lowered with the ``metric'' $h_{ab}={\rm diag}(1,-1)$. Similarly,
the ``propagator'' to be used in Feynman graph expansions is the
full matrix $G^{ab}$, and the interaction terms should be read
out of the CTP classical action $S[\Phi^1]-S[\Phi^2]$, discussed in
Sec. {\bf 2.1}.

In the case of vacuum initial conditions,
these can be included into the path integral by tilting the branches
of the CTP in the complex $t$ plane (the direct branch should acquire an
infinitesimal positive slope, and the return branch, a negative one
 \cite{Mills}).
The CTP boundary condition, that the histories at either branch
should fit continuously at the surface $t=T$, may also be explicitly
incorporated into the path integral as follows. We first include
under the integration sign a term

\begin{equation}
\prod_{x\in R^3}\delta (\Phi^1(x,T)-\Phi^2(x,T))\label{CTPbc1}
\end{equation}

which  enforces this boundary condition;  then  we  rewrite
Eq.(\ref{CTPbc1}) as

\begin{equation}
exp\{(-1/\alpha^2)\int~d^3x~(\Phi^1(x,T)-\Phi^2(x,T))^2\}\nonumber
\end{equation}

where $\alpha\to 0$. This term has the form

\begin{equation}
exp\{i\int~d^4x~d^4x'K_{ab}(x,x')\Phi^a(x)\Phi^b(x')\},\nonumber
\end{equation}

where

\begin{equation}
K_{ab}(x,x')=(i/\alpha^2)\delta
(x-x')\delta(t-T)[2\delta_{ab}-1].\nonumber
\end{equation}

In this way, we have traded the boundary condition by an explicit
coupling to a non local external source.

As before,
variation of the CTP 2PI effective action yields the equations of
motion for background fields and propagators.
The big difference is that now  these equations are real and
causal \cite{87,CalHu88}.

We can now see how the CTP technique solves the ordering problem
in the definition of quantum amplitudes for correlation histories.
One simply considers the specified kernels as products of
fields defined on a closed time - path. In this way, we may
define up to four different kernels $G^{ab}$ independently, to be
identified with the four different possible orderings of the
fields (for simplicity, we assume the background fields are
kept equal to zero). If the kernels $G^{ab}$ can actually be
decomposed as products of c-number fields on the CTP, then
we associate to them the quantum amplitude

\begin{equation}
\Psi [G^{ab}]=\int~D\Phi^a~e^{iS}\prod_{x\gg x',ab}
\delta (\Phi^a(x)\Phi^b(x')-G^{ab}(x,x'))\label{psiofGab}
\end{equation}
(where $S$ stands for the CTP classical action)
The path integral can be manipulated as in Sec. 2.1 to yield

\begin{equation}
\Psi [G^{ab}]\sim [{\rm Det}\{{\partial^2\Gamma\over
\partial G^{ab}\partial G^{cd}}\}]^{(1/2)}
e^{i\Gamma [G^{ab}]}\label{psiofG,2PI,CTP}
\end{equation}
where $\Gamma$ stands now for the CTP 2PI effective action. This last
expression can be analytically extended to more general propagator
quartets, and, indeed, even to kernels which do not satisfy the
relationships $G^{11}(x,x')=G^{21}(x,x')=G^{12*}(x,x')=G^{22*}(x,x')$
for $t\ge t'$, which follow from their interpretation as field
products.

Quantum amplitudes for correlation histories including higher order
products are defined following a similar procedure. For example,
four particle correlations are specified by introducing $16$
kernels \cite{CalHu88}

\begin{equation}
G^{abcd}\sim\Phi^{a}\Phi^{b}\Phi^{c}\Phi^{d}
-G^{ab}G^{cd}
-G^{ac}G^{bd}
-G^{ad}G^{bc}\label{Gabcd}
\end{equation}
If the new kernels are simply products of the binary ones, then
the amplitude is given by
\begin{eqnarray}
\Psi [G^{ab},G^{abcd}]&&=\int~D\Phi^a~e^{iS}\prod_{ab}
\delta (\Phi^a\Phi^b-G^{ab})\nonumber\\
&&\prod_{abcd}\delta (
\Phi^{a}\Phi^{b}\Phi^{c}\Phi^{d}
-G^{ab}G^{cd}
-G^{ac}G^{bd}
-G^{ad}G^{bc}-G^{abcd})\nonumber\\
\label{psiofGabGabcd}
\end{eqnarray}
(In the last two equations, we have included the space - time index $x$
and the branch index $a$ into a single multi index). Here, each pair
appears only once in the product, as well as each quartet $abcd$.
Exponentiating the $\delta$ functions we obtain
\begin{eqnarray}
\Psi [G^{ab},G^{abcd}]=&&\int~DK_{abcd}\int~DK_{ab}\int~D\Phi~
exp\{i[S+{1\over 2}K_{ab}(\Phi^a\Phi^b-G^{ab})\nonumber\\
&&+{1\over 24}K_{abcd}(
\Phi^{a}\Phi^{b}\Phi^{c}\Phi^{d}
-G^{ab}G^{cd}
-G^{ac}G^{bd}
-G^{ad}G^{bc}-G^{abcd})
]\}\nonumber\\
\label{psiofGab,Gabcd,K}
\end{eqnarray}
Now the integral over fields yields the CTP generating functional
for connected graphs, for a theory with a non local interaction
term. Thus
\begin{eqnarray}
\Psi [G^{ab},G^{abcd}]&&=\int~DK_{abcd}\int~DK_{ab}
{}~exp\{i[W[K_{ab},K_{abcd}]-{1\over 2}K_{ab}G^{ab}\nonumber\\
&&-{1\over 24}K_{abcd}(
G^{ab}G^{cd}
+G^{ac}G^{bd}
+G^{ad}G^{bc}+G^{abcd})
]\}\nonumber\\
\label{longequation}
\end{eqnarray}
The integral may be evaluated by saddle point methods, the saddle
being the solution to $W_{,K_{ab}}=(1/2)G^{ab}$,
$W_{,K_{abcd}}=
{1\over 24}(
G^{ab}G^{cd}
+G^{ac}G^{bd}
+G^{ad}G^{bc}+G^{abcd})$. To evaluate the exponential at the saddle
is the same as to perform a Legendre transform on $W$~--it yields
the higher order CTP effective action $\Gamma [G^{ab},G^{abcd}]$.
Variation of $\Gamma$ yields the equation of motion for its
arguments, which are also the inversion of the saddle point
conditions
\begin{eqnarray}
\Gamma_{,G^{ab}}=&&(-1/2)K_{ab}-(1/4)K_{abcd}G^{cd}\nonumber\\
\Gamma_{,G^{abcd}}=&&(-1/24)K_{abcd}\nonumber\\
\label{vareqs}
\end{eqnarray}
Thus up to quartic correlations, the quantum mechanical amplitude is
given by

\begin{equation}
\Psi [G^{ab},G^{abcd}]\sim e^{i\Gamma [G^{ab},G^{abcd}]}\label{finalres}
\end{equation}

This expression can likewise be extended to more general kernels.

As a check on the plausibility of this result, let us note the following
point. Since quantum mechanical amplitudes are additive, it should be
possible to recover our earlier ansatz Eq. (\ref{psiofG,2PI}) for
binary correlation histories from the more general result Eq. (\ref{finalres}),
by integration over the fourth order kernels. Within the saddle
point approximation, integration amounts to substituting these
kernels by the solution to the second Eq. (\ref{vareqs}) for the
given $G^{ab}$, with $K_{abcd}=0$, and with null initial conditions.
(Indeed, since initial conditions can always be included as delta
function - like singularities in the external sources, the third
condition is already included in the second.) This procedure
effectively reduces the fourth order effective action to the 2PI
CTP one \cite{CalHu88}, as we expected.

A basic point which emerges here relevant to our study of decoherence is
that, while quantum field theory is unitary and thus time
reversal invariant, the evolution of the propagators
derived from the 2PI CTP effective action is manifestly
irreversible \cite{CalHu88,89}. The key to this apparent paradox is that, while
the evolution equations are indeed time reversal invariant,
when higher order kernels are retained as independent
variables, their reduction to those
generated by the 2PI effective action involves the
imposition of trivial boundary conditions in the past. Thus the origin of
irreversibility in the two point functions is the same
as in the BBGKY formulation in statistical mechanics \cite{Akhiezer}.
The lesson for us in the present context is that there is an intrinsic
connection between dissipation and decoherence
\cite{HuBelgium,PazSin1}. Knowledge that the evolution of
the propagators  generated by the 2PI effective action
is generally dissipative leads us to expect that
histories defined through binary correlations will
usually decohere. We proceed now to a detailed study of this point.

\newpage
\section{Decoherence of Correlation Histories}

\subsection{Decoherence Functional for Correlation Histories}

Having found an acceptable ansatz for the quantum mechanical
amplitude associated with a correlation history, we are in
a position to study the decoherence functional between two
such histories. As was discussed in the Introduction,
if the decoherence functional is diagonal, then correlation
histories support a consistent probability assignment,
and  may thus be viewed as  classical (stochastic) histories.

For concreteness, we shall consider the simplest case of
decoherence among histories defined through (time-ordered)
binary products. Let us start by considering two histories,
associated with kernels $G(x,x')$ and $G'(x,x')$, which can
in turn be written as products of fields. Taking notice of
the similarity between the quantum amplitudes Eqs. (\ref{psiofalfa})
and (\ref{psiofG}), we can  by analogy to Eq. (\ref{dofalfa})
define the decoherence functional for second correlation order as
\begin{eqnarray}
D[G,G']&&=
\int~d\Phi d\Phi '~
e^{i(S[\Phi ]-S[\Phi '])}\nonumber\\
&&\prod_{x\gg x'}\delta ((\Phi (x)
\Phi (x')-G(x,x'))\delta ((\Phi '(x)
\Phi '(x')-G'^*(x,x'))\nonumber\\
\label{DofGG'}
\end{eqnarray}
Recalling the  expression  Eq.   (\ref{psiofGab}) for the quantum amplitude
associated with the most general binary correlation history, we can rewrite
Eq. (\ref{DofGG'}) as

\begin{equation}
D[G,G']=\int~DG^{12}~DG^{21}~\Psi [G^{11}=G,G^{22}=G'^*,G^{12},G^{21}]
\label{DofGG'2pi}
\end{equation}

This expression for the decoherence functional can be
extended to arbitrary kernels.

In the spirit of our earlier remarks, we use the ansatz
Eq. (\ref{psiofG,2PI,CTP}) for the CTP quantum amplitude and
perform the integration by saddle point methods to obtain

\begin{equation}
D[G,G']\sim e^{i\Gamma [G^{11}=G,G^{22}=G'^*,G^{12}_0,G_0^{21}]}
\label{DofGG'CTP}
\end{equation}
where the Wightman functions are chosen such that

\begin{equation}
{\partial\Gamma\over\partial G^{12}_0}=
{\partial\Gamma\over\partial G^{21}_0}=0
\label{2pisaddle}
\end{equation}
for the given values of the Feynman and Dyson functions.
These last two equations are the sought-for expression for the
decoherence functional.

As an application, let us study the decoherence functional for
Gaussian fluctuations around the vacuum
expectation value (VEV) of the propagators for a $\lambda\Phi^4$
theory, carrying the calculations to two-loop accuracy. Gaussian
fluctuations means that we only need the closed time-path
2PI effective action to second order in the fluctuations
$\delta G^{ab}=G^{ab}-\Delta_0^{ab}$,
where $\Delta_0^{ab}$ stands for the VEVs. Since the effective action
is stationary at the VEV, there is no linear term. Formally
\begin{equation}
\Gamma [\delta G^{ab}]=(1/2)\{\Gamma_{,(aa),(bb)}\delta G^{aa}
\delta G^{bb}+2\Gamma_{,(a\not= b),(cc)}\delta G^{a\not= b}
\delta G^{cc}+\Gamma_{,(a\not= b),(c\not= d)}\delta G^{a\not= b}
\delta G^{c\not= d}\}
\label{John}
\end{equation}
so the saddle point equations (\ref{2pisaddle}) become

\begin{equation}
\{\Gamma_{,(a\not= b),(c\not= d)}\}\delta G_0^{c\not= d}=-
\Gamma_{,(a\not= b),(ee)}\delta G^{ee}
\label{roamingbuffalo}
\end{equation}

The formal Feynman graph expansion of the 2PI effective action
is given in Eq. (\ref{2PIEA}). To two-loop accuracy, we find \cite{CalHu88}
\begin{eqnarray}
\Gamma_2[G^{ab}]&&=-{\lambda\over 8}h_{abcd}\int~d^4x~G^{ab}(x,x)
G^{cd}(x,x)\nonumber\\
&&+{i\lambda^2\over 48}h_{abcd}h_{efgh}\int~d^4x~d^4x'
G^{ae}(x,x')G^{bf}(x,x')G^{cg}(x,x')G^{dh}(x,x')\nonumber\\
\label{88,3.6}
\end{eqnarray}
where $h_{ab},h_{abcd}=1$ if $a=b=c=d=1$, $-1$ if $a=b=c=d=2$, and vanish
otherwise.

Computing the necessary derivatives, we find
\begin{eqnarray}
&&{\partial^2\Gamma\over\partial G^{ab}(x,x')\partial G^{cd}(x'',x''')}
=\nonumber\\
&&({-1\over 2})
[-i (G^{-1})_{ac}(x,x'')(G^{-1})_{db}(x''',x')\nonumber\\
&&+(1/2)\lambda h_{abcd}\delta (x'-x)\delta (x''-x)\delta (x'''-x)
\nonumber\\
&&-(i/2)\lambda^2
h_{aceg}h_{bdfj}\delta (x''-x)\delta (x'''-x')G^{ef}(x,x')G^{gj}(x,x')
]\nonumber\\
\label{92}
\end{eqnarray}
These derivatives are evaluated at $G^{ab}=\Delta_0^{ab}$, where
\begin{eqnarray}
(\Delta_0^{-1})_{ab}(x,x')&&=i[h_{ab}(-\nabla^2+m^2-ih_{ab}\epsilon )\delta
(x'-x)\nonumber\\&&+(\lambda /2)h_{abcd}\delta (x'-x)\Delta_0^{cd}(x,x)
\nonumber\\&&-(i/6)\lambda^2h_{aecd}h_{bfgh}
\Delta_0^{ef}(x,x')\Delta_0^{cg}(x,x')\Delta_0^{dh}(x,x')]\nonumber\\
&&+{1\over 2\alpha^2}\delta (x'-x)\delta (t-T)[2\delta_{ab}-1]\nonumber\\
\label{lasteq}
\end{eqnarray}
where it is understood that the limits $\epsilon ,\alpha\to 0$, $T\to\infty$
are
taken. The first
infinitesimal is included to enforce appropiate Feynman/Dyson
orderings, the  second  to  carry  the  CTP boundary conditions in the far
future.

In computing the Feynman graphs in these expressions, the usual
divergences crop up. They may be regularized and renormalized by
standard methods, which we will not discuss here. The
``tadpole'' graph $\Delta_0^{cd}(x,x)$ can be made to vanish by a
suitable choice of the renormalization point, which we shall assume.

Let us narrow our scope to a physically meaningful set of histories,
namely, those describing
ensembles of real particles distributed with a position-independent
spectrum $f(k)$, $k$ being the four momentum vector. Such
ensembles are described by propagators  \cite{CalHu88}

\begin{equation}
\delta G(x,x')=2\pi\int~({d^4k\over (2\pi )^4})~e^{ik(x-x')}\delta (k^2+m^2)
f(k)\label{ensemble}
\end{equation}

The distribution functions $f$ are real, positive, and even in $k$.
We wish to analyze under what conditions it is possible to assign
consistent probabilities to different spectra $f$. To this end we
must compute the decoherence functional between the propagator in
Eq. (\ref{ensemble}) and another, say, associated with a function $f'$.

Let us begin by investigating Eqs (\ref{roamingbuffalo})
for the missing propagators $G^{12}$ and $G^{21}$. We shall first
disregard the boundary condition enforcing terms in these
equations, introducing them at a later stage. When this is
done, the right hand side of Eqs. (\ref{roamingbuffalo}) vanishes,
since $(-\nabla^2+m^2)G^{aa}(x,x')\equiv 0$ in the present case.

On the other hand, we only need the left hand side to zeroth
order in $\lambda$, since any other term would be of too high an
order to contribute to the decoherence functional at the desired
accuracy. With this in mind, Eq. (\ref{roamingbuffalo})
reduces to the requirement that the unknown propagators should
be homogeneous solutions to the Klein-Gordon equation on both
of their arguments.

To determine the proper boundary conditions for these propagators,
we may consider the boundary terms in Eq.  (\ref{lasteq}), or else
appeal to their physical interpretation. We shall choose the
second approach.

To this end, we observe that
the physical meaning of the propagators as (non standard) products
of fields, Eqs. (\ref{Dick}) to (\ref{dyson}), entails the identity
$G^{12}+G^{21}=G^{11}+G^{22}$, which is consistent in this case,
since both sides solve the Klein - Gordon equation. Actually,
this identity is satisfied by the VEV propagators, so it can
be imposed directly on their variations.

Physically, a change in the propagators reflects a corresponding change
in the statistical state of the field. To zeroth order in the
coupling constant, however, the commutator of two fields is a
c-number , and does not
depend on the state. Therefore, to this accuracy, $G^{12}-G^{21}$
should not change; that is, $\delta G^{12}$ should be equal to
$\delta G^{21}$. We thus conclude that the correct solution to
Eq. (\ref{roamingbuffalo}) is

\begin{equation}
\delta G^{12}=\delta G^{21}=({1\over 2})\{\delta G^{11}+\delta G^{22}\}
\label{wightmen}
\end{equation}
Consideration of the CTP boundary conditions would have led to
the same result.

We may now evaluate the second variation of the 2PI CTP effective
action, Eq. (\ref{John}). We should stress that the Klein-Gordon
operator annihilates all propagators involved, and that
the $O(\lambda )$ term in $\Delta_0^{-1}$ vanishes because
of our choice of renormalization point. Therefore the second (mixed)
term in Eq. (\ref{John}) is of higher than second order
and may be disregarded. The same holds for terms of the form
$(\Delta_0)^{-1}_{ac}\delta G^{cd}(\Delta_0)^{-1}_{db}\delta G^{ab}$,
disregarding boundary terms.

The remaining terms can be read out of Eq. (\ref{92}), with
the input of the ``fish'' graph \cite{Ramond,CalHu88}

\begin{eqnarray}
\Sigma (x,x')&&=(\Delta_0^{11})^2(x,x')=
{i\mu^{\epsilon}\over (4\pi )^2}
\int~{d^4k\over (2\pi )^4}e^{ik(x-x')}
[{2\over\epsilon}
+\ln{m^2\over 4\pi\mu^2}-\psi (1)\nonumber\\
&& - k^2\int_{4m^2}^{\infty}
{d\sigma^2\over\sigma^2(\sigma^2+k^2+i\epsilon )}\sqrt
{1-{4m^2\over\sigma^2}}]\nonumber\\
\label{fish}
\end{eqnarray}
where $\epsilon = d-4$ and $\mu$ is the renormalization scale. Clearly,
the local terms in $\Sigma$ can be absorbed into a coupling-
constant renormalization.

The important thing for us to realize is that the $O(\lambda )$ terms
in Eq. (\ref{92}), as well as the imaginary part of $\Sigma$, contribute only
to the phase of the decoherence functional, and thus are totally
unrelated to decoherence. The only contribution to a
decoherence effect comes from the real part of $\Sigma$. Reading it
out of Eq. (\ref{fish}), we obtain the sought for result

\begin{eqnarray}
\vert D[f,f']\vert&&\sim exp\{({-\pi\lambda^2\over 8})
\int~{d^4p~d^4q\over (2\pi )^8}~
\delta (p^2+m^2)\delta (q^2+m^2)\nonumber\\
&&(f(p)-f'(p))(f(q)-f'(q))
\theta [-((p+q)^2+4m^2)]\sqrt{1+{4m^2\over (p+q)^2}}\}\nonumber\\
\label{juanpa}
\end{eqnarray}
where $\theta$ is the usual step function.
As expected,  we  do  find  decoherence  between  different correlation
histories. Moreover, decoherence is related to dissipative processes,
which in this case arise from pair production \cite{89}.
Indeed, the real part of the kernel $\Sigma$ is
essentially the probability of a real pair being produced out of
quanta with momenta $p$ and $q$, with $p+q=k$ \cite{Ramond}.

Let us mention two obvious consequences of our result for the
decoherence functional. The first point is that decoherence is associated
with instability of the vacuum: the distribution functions whose overlap
is suppressed represent ensembles which are unstable against
non trivial scattering of the constituent particles. This
scattering produces correlations between particles. Therefore,
truncation of the correlation hierarchy leads to an explicitly
dissipative evolution. This would not be the case if there were no
scattering.

The second point is that $\vert D\vert$ remains unity on the diagonal.
Thus, at least
for Gaussian fluctuations, and to two-loop accuracy, all histories
are equally likely. What this means physically is that the
two-point functions to be perceived by an observer after the
quantum to classical transition need not be close to their
VEV in any stringent sense. Indeed, what is observed will
not even be ``vacuum fluctuations'' in the proper sense
of the word; they are real
physical particles whose momenta are on shell, and
may propagate to the asymptotic region, if they manage not to collide
with other particles.

\subsection{Beyond Coarse Graining}

For the observer confined to a single consistent history, as
is the case for the quantum cosmologist, questioning
the probability distribution of histories is somewhat academic.
What would be relevant is one's ability to
predict the future behavior of one's particular history.
This ability is impaired by the lack of knowledge about the
coarse-grained elements of the theory, which, in our case, are the higher
correlations of the field.

As we have already seen, variation of the 2PI CTP effective action,
id est, of the phase of the decoherence functional, yields the
evolution equations for the VEVs of the two-point functions.
These equations should be regarded as the Hartree-Fock
approximation to the actual evolution, since in them the effect
of higher correlations is represented only in the average.
Deviations of the actual evolution from this ideal average may
be represented by adding a source term to the Hartree-Fock equation.
As the detailed state of the
higher correlations is unknown, this right hand side should take
the form of a stochastic binary external source.

The non-diagonal terms of the decoherence functional
represented in Eq. (\ref{juanpa}), while not
contributing to the Hartree-Fock equations, contain the necessary
information to build a phenomenological model of the back reaction
of the higher correlations on the relevant sector. To build this
model, we compare the actual form of the decoherence functional
against that resulting from the coupling of the propagators to
an actual gaussian random external source \cite{Feynman}.

The result of this comparison is that higher correlations
react on the propagators as if these obey a
Langevin- type equation

\begin{equation}
{\partial\Gamma [\delta G^{11}=\delta G,\delta G^{22}=\delta G'^*,
\delta G^{12}_0,\delta G_0^{21}]\over
\partial (\delta G(x,x'))}={-1\over 2v}F(x-x')J(x-x')\label{Langeq}
\end{equation}
where, after the variational derivative is taken, we must take the limit
$\delta G'\to\delta G$. In Eq. (\ref{Langeq})
$v$ is (formally) ``the space - time volume'', the
gaussian stochastic
source $J$ has autocorrelation $\langle J(u)J(u')\rangle
=\delta (u-u')$, and

\begin{equation}
F^2(u)=\lambda^2\int_0^{\infty}{ds\over (4\pi s)^2}
\int_{4m^2}^{\infty}~d\sigma^2~\sin
(s\sigma^2-{u^2\over 4s})\sqrt
{1-{4m^2\over\sigma^2}}\label{theend}
\end{equation}

Because the limit $\delta G'\to\delta G$ is taken, the imaginary terms
of the CTP effective action reproduced in Eq. (\ref{juanpa}) do not
contribute to the  left hand side of Eq. (\ref{Langeq}); as far as the
``Hartree - Fock'' equations are concerned, they could as well be deleted
from the effective action.

However, the stochastic source
in the right hand side of Eq. (\ref{Langeq}) modifies the quantum amplitude
associated with the correlation history by a factor

\begin{equation}
{\rm exp}\{ (i/2)\int~d^4u~F(u)J(u)\delta G(u)\},\nonumber
\end{equation}

 where
$G(u)=(1/v)\int~d^4X~G(X+(u/2),X-(u/2))$. Correspondingly, the
decoherence functional gains a factor

\begin{equation}
{\rm exp}\{ (i/2)\int~d^4u~F(u)J(u)(\delta G(u)-\delta G'(u))\}.\nonumber
\end{equation}

Upon averaging over all possible external sources, each having
a probability

\begin{equation}
{\rm exp}\{ (-1/2)\int~d^4u~J^2(u)\},\nonumber
\end{equation}

 the new
factor in the decoherence functional becomes

\begin{equation}
{\rm       exp}\{        (-1/8)\int~d^4u~F^2(u)(\delta        G(u)-\delta
G'(u))^2\},\nonumber
\end{equation}

which exactly reproduces Eq. (\ref{juanpa}).
Observe that the assumed form for the right hand side of Eq. (\ref
{Langeq}), and the requirement of recovering Eq. (\ref{juanpa})
upon averaging, uniquely determines the function $F$.

In this way,  Eq.(\ref{Langeq})  yields the  correct,  if only a
phenomenological, description of the dynamics of classical
fluctuations in the aftermath of the quantum to classical transition.
It should be obvious that nonlinearity is essential to the generation
of these fluctuations.

\section{Discussion}
This paper presents three main results. The first is the ansatz
Eq. (\ref{psiofG,2PI}) for the quantum amplitude associated with a
correlation history. The second is the ansatz Eq.
(\ref{DofGG'CTP}) for the decoherence functional between two
such histories. On the basis of this ansatz, we have shown in
Eq. (\ref{juanpa}) that the quantum interference between
histories corresponding to different particle spectra is
suppressed whenever these spectra differ by particles whose
added momenta go above the two particle treshold $4m^2$,
$m^2$ being the one-loop radiative-corrected physical mass. The third result
is the phenomenological description in Eq. (\ref{Langeq})
of the dynamics of an individual consistent correlation history.

What we have presented in the above, despite its embryonic form,
is a framework for bringing together the correlational-hierarchy idea
in non-equilibrium statistical mechanics
and the consistent-history interpretation of quantum
mechanics. This framework puts decoherence and  dissipation due to fluctuations
and noise (manifested here through particle creation) on the same footing. It
suggests a natural (intrinsic) measure of coarse-graining which is
commensurate with ordinary accounts of dissipative phenomena, and with it
addresses the issue of quantum to classical transition. It also provides a
theoretical basis for the derivation of classical stochastic equations
from quantum fluctuations, and identifies the nature of noise in these
equations.

It should be noticed that a formal identity exists between the
present results and those previously obtained from the influence
functional formalism \cite{IF,CalLeg83,ZhangPhD}.
Indeed, our decoherence functional has
the same structure as the influence  functional, with the
non diagonal terms in Eq. (\ref{juanpa}) playing the role
of the ``noise kernel''. This is more than an analogy, as it
should be clear from the discussions above and elsewhere.

While for reasons of clarity and economy of space, we have focused on
a simple application from quantum field theory to
develop our arguments, the implications on quantum mechanics and statistical
mechanics go beyond what this example can show.
The theoretical issues raised here in the context of quantum mechanics
and statistical mechanics, as well as the consequences of problems raised
in the context of quantum and semiclassical
(especially the inflationary universe) cosmology,  which
motivated us to make these inquiries in the first place,
will be explored in greater detail elsewhere.

This work is part of an on-going program which draws on many year's
worth of pondering on the role of statistical mechanics ideas
in quantum cosmology, using quantum field theoretical methods
while placing the issues in the larger context of general physics.
The project began in 1985, when one of us (EC) was invited by Dieter Brill
to join the General Relativity Group at Maryland.
It is therefore an honour and a pleasure for us to
dedicate this paper to him on this happy occasion.

\noindent{\bf Acknowledgments}

E. C. is partially supported by the Directorate General for
Science Research and Development of the Commission of the
European Communities under Contract N$^{\rm o}$ C11 - 0540 -
M(TT), and by CONICET, UBA and Fundaci\'on Antorchas
(Argentina). B. L. H's research
is supported in part by the US NSF under grant PHY91-19726
This collaboration is partially supported by NSF and CONICET as part of
the Scientific and Technological Exchange Program between Argentina
and the USA.


\begin{thebibliography}{999}

\bibitem{manyworld}
H. Everitt, III, Rev. Mod. Phys. {\bf 29}, 454 (1957);
B. S. DeWitt and N. Graham, eds.,{\it The Many-Worlds Interpretation of Quantum
   Mechanics}
(Princeton Univ., Princeton, 1973).

\bibitem{conshis}
R. B. Griffiths, J. Stat. Phys. {\bf 36}, 219 (1984);
R. Omn\'es, J. Stat Phys. {\bf 53}, 893, 933, 957 (1988);
Ann. Phys. (NY) {\bf 201}, 354 (1990); Rev. Mod. Phys. {\bf 64}, 339 (1992)

\bibitem{HarMis}
J. B. Hartle, ``Quantum Mechanics of Closed Systems''
in {\it Directions in General Relativity} Vol. 1,
eds B. L. Hu, M. P. Ryan and C. V. Vishveswara
(Cambridge Univ., Cambridge, 1993)

\bibitem{decohis}
M. Gell-Mann and J. B. Hartle, in
{\it Complexity, Entropy and the Physics of Information}, ed.
by W. H. Zurek (Addison-Wesley, Reading, 1990); Phys. Rev. {\bf D47}, (1993)
H. F. Dowker and J. J. Halliwell, Phys. Rev. {\bf D46}, 1580 (1992);
Brun, Phys. Rev. {\bf D47},  (1993)


\bibitem{Akhiezer} A. I. Akhiezer and S. V. Peletminsky, {\it Methods
of Statistical Physics} (Pergamon, London, 1981).

\bibitem{Prigogine} I. Prigogine, {\it Non Equilibrium Statistical
Mechanics} (John Wiley, New York, 1962);
R. Balescu, {\it Equilibrium and Non Equilibrium
Statistical Mechanics} (John Wiley, New York, 1975).

\bibitem{KadBay} L. Kadanoff and G. Baym, {\it Quantum Statistical
Mechanics} (Benjamin, New York, 1962).

\bibitem{KuboTH}  R. Kubo, M. Toda and N. Hashitsume {\it Statistical
        Physics II}, (Springer-Verlag, Berlin, 1978);
J. A. McLennan, {\it Introduction to Non-Equilibrium
Statistical Mechanics} (Prentice-Hall, New Jersey, 1989);
N. G. van Kampen, {\it Stochastic Processes in
Physics and Chemistry} (North Holland, Amsterdam, 1981).


\bibitem{CalHu88} E. Calzetta and B. L. Hu, Phys. Rev. {\bf D37}, 2878
(1988).

\bibitem{CHH88} E. Calzetta, S. Habib and B. L. Hu, Phys. Rev.
{\bf D37}, 2901 (1988);
S. Habib, Ph. D. Thesis, University of Maryland, 1988
(unpublished).


\bibitem{ZhangPhD} Yuhong Zhang, Ph. D. Thesis, University of Maryland,
1990 (unpublished);
B. L. Hu, J. P. Paz and Y. Zhang, Phys. Rev. {\bf D45}, 2843 (1992);
``Quantum Brownian Motion in a General Environment
II.  Nonlinear coupling and perturbative approach'' Phys.  Rev.  {\bf D47},
(1993);
``Stochastic Dynamics of
Interacting Quantum Fields'' (paper III).


\bibitem{CalHuKT2BM}
E. Calzetta and B. L. Hu, ``From Kinetic Theory to Brownian Motion'' (1993);
``Quantum and Classical Fluctuations'' (1993);
``On Correlational Noise'' (1993)

\bibitem{HuSpain} B. L. Hu, ``Fluctuation, Dissipation and Irreversibility
in Cosmology''
in {\it The Physical Origin of Time-Asymmetry} Huelva, Spain, 1991
eds. J. J. Halliwell, J. Perez-Mercader and W. H. Zurek (Cambridge University
Press, 1993).

\bibitem{HuWaseda} B. L. Hu, ``Quantum Statistical Processes in the Early
Universe''
in {\it Quantum Physics and the Universe}, Proc. Waseda Conference, Aug. 1992
ed. Namiki, K. Maeda, et al  (Pergamon Press, Tokyo, 1993).


\bibitem{HuErice} B. L. Hu ``Quantum and Statistical Effects in Superspace
Cosmology''
in {\it Quantum Mechanics in Curved Spacetime}, ed. J. Audretsch
and V. de Sabbata (Plenum, London 1990).

\bibitem{BalVen} R. Balian and M. Veneroni, Ann. Phys. (N. Y.)
{\bf 174}, 229 (1987)

\bibitem{Zurek}  W. H. Zurek, Phys. Rev. {\bf D24}, 1516 (1981);
{\bf D26}, 1862 (1982);
in {\it Frontiers of Nonequilibrium Statistical Physics}, ed. G. T. Moore
and M. O. Scully (Plenum, N. Y., 1986);
W. G. Unruh and W. H. Zurek, Phys. Rev. {\bf D40}, 1071 (1989);
Physics Today {\bf 44}, 36 (1991);
W. H. Zurek, J. P. Paz and S. Habib, Phys. Rev. {\bf  47}, 488 (1993).

\bibitem{Zeh}  E. Joos and H. D. Zeh, Z. Phys. {\bf B59}, 223 (1985);
H. D. Zeh, Phys. Lett. A {\bf 116}, 9 (1986).

\bibitem{WheZur} J. A. Wheeler and W. H. Zurek, {\it Quantum Theory and
Measurement} (Princeton Univ., Princeton, 1983)

\bibitem{CalLeg83} A. O. Caldeira and A. J. Leggett,
Physica {\bf 121A}, 587 (1983);
Phys. Rev. {\bf A31},
1059 (1985);
A. J. Leggett, S. Chakravrty, A. T. Dorsey,
 M.  P.  A.  Fisher, A.  Garg and W.  Zwerger, Rev.  Mod.  Phys.
 {\bf 59}, 1 (1987).

\bibitem{HuTsukuba} B. L. Hu, ``Statistical Mechanics and Quantum Cosmology'',
in {\it Proc. Second International Workshop on Thermal Fields and Their
Applications}, eds. H. Ezawa et al (North-Holland, Amsterdam, 1991);
E. Calzetta, Phys. Rev. {\bf D43}, 2498 (1991);
S. Sinha, Ph. D. Thesis, University of Maryland,
1991 (unpublished);
S. Sinha and B. L. Hu, Phys. Rev. {\bf D44}, 1028 (1991);
B. L. Hu, J. P. Paz and S. Sinha, ``Minisuperspace as a Quantum  Open System''
in {\it Directions in General Relativity}  Vol. 1, (Misner Festschrift)
eds B. L. Hu, M. P. Ryan and C. V. Vishveswara
(Cambridge Univ., Cambridge, 1993).

\bibitem{HalGR13} J. J. Halliwell,
in Proc. 13th GRG Meeting, Cordoba, Argentina, July 1992

\bibitem{PazSin1} J. P. Paz and S. Sinha, Phys. Rev. {\bf D44}, 1038 (1991);
{\it ibid} {\bf D45}, 2823 (1992);
E. Calzetta and D. Mazzitelli, Phys. Rev. {\bf D42},
4066 (1990).

\bibitem{HuPhysica} B. L. Hu, Physica A {\bf 158}, 399 (1989).

\bibitem{CalCQG} E. Calzetta, Class. Quan. Grav. {\bf 6}, L227 (1989)

\bibitem{HuBelgium} B. L. Hu, J. P. Paz and Y. Zhang, ``Quantum Origin of
Noise and Fluctuation in Cosmology'' in {\it Proc. Conference on
the Origin of Structure in the Universe} Chateau du Pont d'Oye, Belgium,
April, 1992, ed. E. Gunzig and P. Nardone (NATO ASI Series) (Plenum Press,
New York, 1993); S. Habib and H. E. Kandrup, Phys. Rev. {\bf D46}, 5303 (1992).

\bibitem{HuZhaDrexel} B. L. Hu and Y. Zhang, in
{\it Proc. Third International
Workshop on Quantum Nonintegrability}, Drexel University, Philadelphia,
May 1992, ed. D. H. Feng, J. Yuan (Gordon and Breach, New York, 1993)

\bibitem{correlation}
R. Geroch,  No\^{u}s {\bf 18}, 617 (1984);
J.B Hartle, in {\em Gravitation in Astrophysics}, 1986
NATO Advanced Summer Institute, Cargese, ed. B.Carter and J. Hartle
(NATO ASI Series B: Physics Vol. 156, Plenum, N. Y., 1987)

\bibitem{HarBri} J. B. Hartle,
in {\it Directions in General Relativity} Vol 2 (Brill Festschrift)
eds. B. L. Hu and T. A. Jacobson (Cambridge Univ., Cambridge, 1993)

\bibitem{Penrose} R. Penrose, in Proc. 13th GRG Meeting, Cordoba, Argentina,
July, 1992.

\bibitem{inflation}  A. H. Guth, Phys. Rev. {\bf D23}, 347 (1981);
 K. Sato, Phys. Lett. {\bf 99B}, 66 (1981);
 A. D. Linde, Phys. Lett. {\bf 108B}, 389 (1982);
 A. Albrecht and P. J. Steinhardt, Phys. Rev. Lett. {\bf 48}, 1220 (1982);
Also see E. Kolb and M. Turner {\it the Early Universe} (Addison-Wesley,
Menlo Park, 1990), and A. Linde, {\it Inflationary and Quantum Cosmology}
(Academic Press, San Diego, 1991)

\bibitem{QC} J. B. Hartle and S. W. Hawking, Phys. Rev. {\bf D28}, 1960 (1983);
 A. Vilenkin, Phys. Lett. {\bf 117B}, 25 (1985); Phys. Rev. {\bf D27},
2848 (1983), {\bf D30}, 509 (1984);


\bibitem{Nakajima} S. Nakajima, Progr. Theor. Phys. {\bf 20}, 948 (1958);
R. Zwanzig, J. Chem. Phys. {\bf 33}, 1338 (1960), and
in {\it Lectures in Theoretical
Physics III}, edited by W. Britten {\it et al.}(Wiley, New York, 1961),
p. 106;
H. Mori, Prog. Theor. Phys. {\bf 34}, 423 (1965)

\bibitem{HarSTCG} J. B. Hartle,  Phys.  Rev.  {\bf D37}, 2818 (1988),

\bibitem{HalSOH}J. J.  Halliwell and M. Ortiz, ``Sum over
Histories Origin of the Composition Laws of Relativistic Quantum Mechanics''
Phys. Rev. {\bf D47} (1993).


\bibitem{StoInf} A. A. Starobinsky,
in {\it Field Theory, Quantum Gravity and
Strings}, ed. H. J. de Vega and N. Sanchez (Springer, Berlin 1986);
J. M. Bardeen and G. J. Bublik, Class. Quan. Grav. {\bf 4}, 473 (1987).

\bibitem{HuZha90} B. L. Hu and Y. Zhang,
``Coarse-Graining, Scaling, and Inflation''
Univ. Maryland Preprint 90-186;
B. L. Hu, in {\it Relativity and Gravitation: Classical
and Quantum} Proc. SILARG VII, Cocoyoc, Mexico 1990.
eds. J. C. D' Olivo et al (World Scientific, Singapore 1991);
S. Habib, Phys. Rev. {\bf D46}, 2408 (1992).

\bibitem{Wigner}  E. P. Wigner, Phys. Rev. {\bf 40}, 749 (1932)


\bibitem{SchKel} J. Schwinger, J. Math. Phys. {\bf 2} (1961) 407;
L. V. Keldysh, Zh. Eksp. Teor. Fiz. {\bf 47 }, 1515 (1964)
[Engl. trans. Sov. Phys. JEPT {\bf 20}, 1018 (1965)].

\bibitem{Zhou}G. Zhou, Z. Su, B. Hao and L. Yu, Phys. Rep. {\bf 118},
1 (1985);
Z. Su, L. Y. Chen, X. Yu and K. Chou, Phys. Rev. {\bf B37}, 9810 (1988).

\bibitem{DeWJor} B. S. DeWitt, in {\it Quantum Concepts in Space and Time}
ed. R. Penrose and C. J. Isham (Claredon Press, Oxford, 1986);
R. D. Jordan, Phys. Rev. {\bf D33}, 44 (1986).

\bibitem{87} E. Calzetta and B. L. Hu, Phys.  Rev.  {\bf D35}, 495
(1987).

\bibitem{Ramond} P. Ramond, {\it Field Theory, a Modern Primer}
(Benjamin, New York, 1981).

\bibitem{Jackiw}R. Jackiw, Phys. Rev. {\bf D9}, 1686 (1974);
J. Iliopoulos, C. Itzykson and A. Martin, Rev. Mod. Phys. {\bf 47},
165 (1975).

\bibitem{Langer}J. S. Langer, Ann. Phys. (NY) {\bf 41}, 108 (1967);
{\bf 54}, 258 (1969).

\bibitem{2PI}H. D. Dahmen and G. Jona - Lasinio, Nuovo Cimento
{\bf 52A}, 807 (1962); C. de Dominicis and P. Martin, J. Math. Phys.
{\bf 5}, 14 (1964); J. M. Cornwall, R. Jackiw and E. Tomboulis,
Phys. Rev. {\bf D10}, 2428 (1974); R. E. Norton and J. M. Cornwall,
Ann. Phys. (NY) {\bf 91}, 106 (1975).

\bibitem{Spino} E. Calzetta, Ann. Phys. (NY), {\bf 190}, 32 (1989).

\bibitem{Mills} R. Mills, {\it Propagators for Many Particle Systems}
(Gordon and Breach, New York, 1969).

\bibitem{89} E. Calzetta and B. L. Hu, Phys.  Rev.  {\bf D40}, 656
(1989).

\bibitem{Feynman} Our argument is adapted from a similar problem in
R. Feynman and A. Hibbs, {\it Quantum Mechanics and Path Integrals},
(McGraw - Hill, New York, 1965).

\bibitem{IF} R. Feynman and F. Vernon, Ann. Phys. (NY) {\bf 24},
118 (1963).


\end{thebibliography}
\end{document}